\documentclass[9pt,twocolumn,twoside]{osajnl}

\journal{pr} 

\setboolean{shortarticle}{false}

\title{Visualization of magnetic fields with cylindrical vector beams in a warm atomic vapor}

\author[1]{Shuwei Qiu}
\author[1,2]{Jinwen Wang}
\author[2]{Francesco Castellucci}
\author[3,+]{Mingtao Cao}
\author[3]{Shougang Zhang}
\author[4]{Thomas W. Clark}
\author[2]{Sonja Franke-Arnold}
\author[1,*]{Hong Gao}
\author[1]{Fuli Li}

\affil[1]{Ministry of Education Key Laboratory for Nonequilibrium Synthesis and Modulation of Condensed Matter, Shaanxi Province Key Laboratory of Quantum Information and Quantum Optoelectronic Devices,
School of Physics, Xi'an Jiaotong University, Xi'an 710049, China}
\affil[2]{School of Physics and Astronomy, University of Glasgow, Glasgow G12 8QQ, United Kingdom}
\affil[3]{Key Laboratory of Time and Frequency Primary Standards, National Time Service Center, Chinese Academy of Science, Xi'an, 710600, China}
\affil[4]{Institute for Solid State Physics and Optics, Wigner Research Centre for Physics, H-1525 Budapest P.O. Box 49, Hungary}

\affil[+]{Corresponding author: mingtaocao@ntsc.ac.cn}
\affil[*]{Corresponding author: honggao@xjtu.edu.cn}

\dates{Compiled \today}

\begin{abstract}
We propose and demonstrate an experimental implementation for the observation of magnetic fields from spatial features of absorption profiles in a warm atomic vapor. A radially polarized vector beam that traverses an atomic vapor will generate an absorption pattern with petal-like structure by the mediation of a transverse magnetic field (TMF). The spatial absorption pattern rotates when the azimuthal angle of the TMF is changed, while its contrast decreases when the longitudinal component of the magnetic field increases. By analyzing the intensity distribution of the transmitted pattern we can determine the magnetic field strength. Our work provides a framework for investigating three-dimensional magnetic field distributions based on atoms.  
\end{abstract}

\setboolean{displaycopyright}{true}

\begin{document}

\maketitle

\section{Introduction}
Perhaps the most remarkable demonstrations of coherent interaction between atoms and photons are electromagnetically induced transparency (EIT), electromagnetically induced absorption (EIA) and coherent population trapping  \cite{boller1991observation,fulton1995effects,alezama1999eia,renzoni1997coherent}. These processes can be interpreted as a consequence of quantum interference -- they are based on the fact that an optical field can transform atomic states such that an atomic transition can be entirely suppressed and subsequent absorption eliminated. Quantum interference shows an exceptional sensitivity to frequency shifts, including those induced by magnetic fields. This makes atomic ensembles excellent tool for magnetometry, with potential application across research fields as diverse as biomedicine, seismology, defense, and general metrology \cite{kitching2011atomic,wiesendanger2011single}. 

Atomic magnetometers can now reach excellent sensitivites, comparable to, and even surpassing those of superconducting quantum interference devices (SQUIDs) \cite{drung1990low,pannetier2004femtotesla,kominis2003subfemtotesla}.  Since the first demonstration of EIT-based scalar magnetometers \cite{fleischhauer1994quantum}, various schemes have been reported, relying on the zero field resonance observed in Hanle-type experiments \cite{alipieva2003narrow,gateva2007shape}, on optical pumping \cite{acosta2006nonlinear,afach2015highly,bison2018sensitive,zhang2019multi}, or the nonlinear Faraday effect in a manifold of a single ground state \cite{novikova2001compensation,pustelny2006pump,budker1998nonlinear,budker2000sensitive,novikova2000ac,pustelny2008magnetometry}. Miniaturisation presents a challenge for atomic magnetometers, but over the last two decades devices have been developed that combine extreme sensitivity with minute detection volumes \cite{kominis2003subfemtotesla,shah2007subpicotesla}.

Most atomic magnetometers perform scalar magnetic field metrology, i.e.~determined the magnetic field along a pre-defined axis. Simultaneously measuring the strength and direction of a magnetic field would be of great importance in specific areas such as satellite navigation and biological magnetic field measurement \cite{budker2007optical,le2013optical}.
The direction of a magnetic field can be addressed by vector magnetometers, first demonstrated in \cite{lee1998sensitive}. Since then, various schemes characterized by EIT or its counterpart, EIA, have been extensively studied \cite{dimitrijevic2008role,yudin2010vector,cox2011measurements,margalit2013degenerate}. The full vector nature of a magnetic field may be accessed by simultaneously probing the magnetic field in orthogonal directions by separate probe beams.  Alternatively, adding an external transverse magnetic field (TMF) can make EIT-based methods sensitive to different magnetic field components by considering polarization rotation or resonance amplitudes \cite{yudin2010vector,cox2011measurements}. 

In this work we explore the possibility to detect both the strength and alignment of a magnetic vector field from the interaction of a warm atomic vapor with a vector beam (VB), i.e.~a light field that has a polarisation pattern that is varying across the beam profile.    
The interaction of vector beams with atoms is a relatively new concept \cite{zhan2009cylindrical,wang2020vectorial}, which has been used to explore spatial anisotropy \cite{fatemi2011cylindrical,wang2018optically,yang2019manipulating,wang2019directly,wang2020optically}, nonlinear effects \cite{shi2015magnetic,stern2016controlling,bouchard2016polarization,hu2019nonlinear} and quantum storage \cite{parigi2015storage,ye2019experimental}. 

Of particular interest to this work is the extension of EIT, conventionally observed as spectral features with homogeneously polarised probe beams, to spatially resolved EIT resulting from inhomogeneously polarised VBs.  This effect has been observed both in cold \cite{radwell2015spatially} and warm \cite{yang2019observing} atomic systems. In the former case, a weak TMF closes the EIT transitions, thereby generating phase-dependent dark states and, in turn, spatially dependent transparency. As the spatially observed transparency patterns and applied magnetic fields are directly coupled, this offers the possibility of detecting magnetic fields from absorption profiles  \cite{clark2016sculpting,castellucci2021atomic}.

In this paper, an experimental setup is presented to visually observe the magnetic field based on Hanle resonances in a warm atomic vapor. Importantly, we analyze spatially resolved absorption patterns instead of the time-resolved spectrum, which is fundamentally different from other aforementioned methods.
By employing VBs, we show that the absorption pattern is sensitive to the TMF strength, visible particularly in the degree of absorption, whereas maximal transmission remains unchanged. It's worth noting that the above results will change depending on experimental parameters, \textit{e.g.} increasing the temperature of the gas should lead to a reduction of transparency throughout the whole beam profile. Furthermore, the transmitted pattern of VBs can be rotated arbitrarily according to the alignment of TMF. For the general case in the current work, the spatial magnetic field can be decomposed into a TMF and a longitudinal magnetic field (LMF) according to the quantization axis. The absorption patterns and corresponding polar plots can then be analyzed to recover the full magnetic field information. Such a procedure could prove to be a powerful tool to measure the three-dimensional (3D) magnetic distribution, and can even be applied in room temperature atomic vapors, simplifying future atomic magnetometer design.

\begin{figure}[htbp]
\centering
\includegraphics[width=\linewidth]{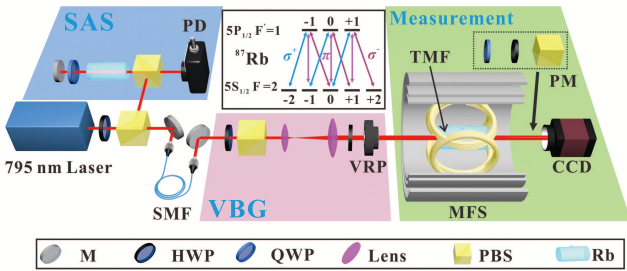}
\caption{The schematic of the experimental setup and atomic energy levels. M: mirror; HWP: half-wave plate; QWP: quarter-wave plate; L: lens; PBS: polarization beam splitter; PD: photodetector; SMF: single mode fiber; VRP: vortex retarder plate; CCD: charge-coupled device camera; MFS: Magnetic field shielding; PM: Projection measurement. SAS: Saturated absorption spectroscopy; VBG: Vector beam generation.}
\label{fig1}
\end{figure} 

\section{Experimental setup}
The experimental setup is shown in Fig. \ref{fig1}. The output of a frequency locked 795 nm external cavity diode laser is sent through a single-mode fiber (SMF) to improve the mode quality of the Gaussian beam. 
After the fiber, the beam passes through a half-wave plate and a polarizing beam splitter (PBS) to adjust the beam intensity and fix the polarized state of the beam as horizontal polarization. A telescope is applied to expand the beam size and the achieved high-quality Gaussian beam waist is 4 mm. The VBs are generated by sending the linearly polarized beam through a vortex retarder plate (VRP), a liquid-crystal-based retardation wave plate with an inhomogeneous optical axis which displays an azimuthal topological charge \cite{marrucci2006optical,marrucci2006pancharatnam}. 
The laser frequency is locked to the $5{S_{{1 \mathord{\left/
 {\vphantom {1 2}} \right.
 \kern-\nulldelimiterspace} 2}}},F = 2 \to 5{P_{{1 \mathord{\left/
 {\vphantom {1 2}} \right.
 \kern-\nulldelimiterspace} 2}}},F' = 1$ transition of the $^{87}$Rb D1 line.
The Rb cell has a length of 50 mm. A three-layer $\mu$-metal magnetic shield is used to isolate the atoms from the environmental magnetic fields. The temperature of the cell is set at 60°C with a temperature controller. A solenoid coil (not shown in figure) inside the inner layer generates a uniform LMF, oriented along the light propagation direction, $\mathbf{k}$. A TMF, in the plane perpendicular to $\mathbf{k}$ and coverin the whole cell, is generated by two pairs of orthogonal Helmholtz coils, each pair independently controlled by a high precision current supply driver.  By adjusting the current ratio, and hence the horizontal and vertical TMF component, it is possible to produce an arbitrary TMF. The power of the incident laser beam is 3.4 $\rm mW/cm^{2}$ ( $\approx$ 0.75 $I_{\rm sat}$). After passing through the cell, the spatial intensity distribution of the beam is recorded by a charge-coupled device camera (CCD). 

\begin{figure}[ht!]
\centering
\includegraphics[width=\linewidth]{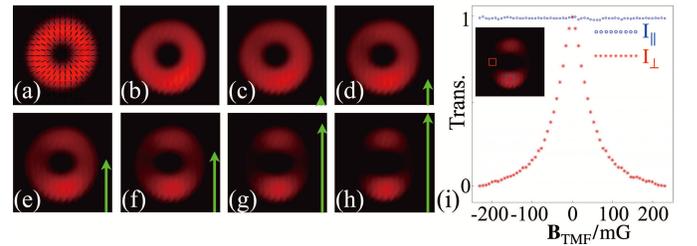}
\caption{The experimental results of the radially polarized beam in presence of TMF. (a) Intensity and polarization distribution without atoms. (b) - (h) Intensity distributions after passing through atoms under vertical TMF of varying strength: ${\rm B}_{\rm TMF}$ = 0 mG, 23 mG, 61 mG, 123 mG, 146 mG, 206 mG and 230 mG, respectively. (i) The dependence of transmitted intensity for two selected regions against ${\rm B}_{\rm TMF}$.}
\label{fig2}
\end{figure}

The polarization distribution of the probe VBs can be reconstructed by measuring the Stokes parameters, which represent the full polarization information of  the light \cite{milione2011higher}. Experimentally, the Stokes parameters can be obtained by using the projection measurement system consisting of a QWP, a polarizer and a CCD. Fig. \ref{fig2} (a) shows the polarization distribution of the generated VB with $m=$1, which is also known as a radially polarized beam. Here, $m$ is the polarization topological charge of the VB. It can be seen that this distribution varies periodically with the azimuthal angle in the plane of the beam. The electric field vector of the VBs considered here can be expressed as:
\begin{equation}
\mathbf{E}(r, \phi,z)={E_0}(r, \phi,z)
\begin{pmatrix}
 \cos(m\phi)\\
 \sin(m\phi)\\
 0
\end{pmatrix}.
\label{1}
\end{equation}
Here $r$ is the radial distance, $\phi$ denotes the azimuthal angle and $z$ is the propagation distance. The position-dependent complex amplitude of the light is ${E_0}(r, \phi,z)$, where $m$ is an integer.

\section{Experimental results}

Now we turn to how the magnetic field influences the interaction between the vector beam and the atoms: the transmission in particular. Firstly, when the magnetic field is not applied, there is very little absorption of the vector beam, as compared to the profile without atoms (Fig. \ref{fig2} (a) and (b)). As a vertical TMF is applied to the atoms however, a petal-like transmission pattern gradually appears, as shown in Fig. \ref{fig2} (c)-(h). In general, one predicts $2\times m$ petals, with the exception of 
$m=0$. In our case, we consider $m=1$, and so we observe a two-fold symmetry, as considered in more detail below.

Increasing the magnitude of the TMF, we observe that maximum transmission always occurs in the region where the linear polarization is parallel or antiparallel to the TMF. The strongest absorption occurs in regions where the local linear polarization is perpendicular to the TMF axis, and it increases with increasing magnitude of the TMF. We note that positive and negative TMFs both lead to the same pattern, as atomic transitions respond to the alignment but not orientation of linear polarization. The variation of transparency  and absorption  are plotted against the TMF strength (from $-230$ mG to $+230$ mG) in Fig. \ref{fig2} (i). To make a systematic comparison, a point of maximum transmission and absorption respectively is chosen, and the corresponding probe intensity ($\rm{I_{\parallel}}$ and $\rm{I_{\perp}}$) is determined, averaged over a square area of 25 pixels, to reduce experimental error. The red curve then shows the local sensitivity of the VBs to the TMF strength, in agreement with the Hanle-EIT profile. 

\begin{figure}[ht!]
\centering
\includegraphics[width=\linewidth]{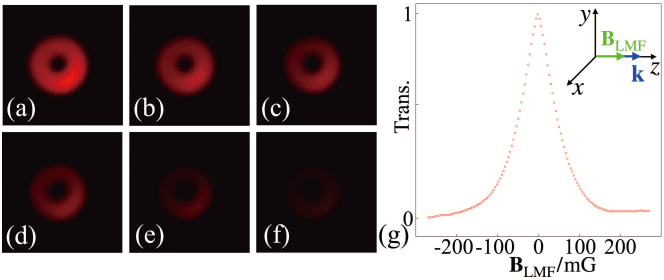}
\caption{The experimental results of the radially polarized beam in presence of an LMF. (a) - (f): the intensity distributions after passing through the atom vapor under the varied LMF: ${\rm B}_{\rm LMF}$ = 0 mG, 50 mG, 100 mG, 120 mG, 160 mG and 200 mG, respectively. (g) The dependence of transmitted intensity for whole beam against the ${\rm B}_{\rm LMF}$.}
\label{fig3}
\end{figure}

As expected, the absorption of the optical ${\mathbf E}$-field components aligned with the $\mathbf B$-field orientation is independent of the magnitude of the applied magnetic field strength. We can therefore monitor these spatial positions to consider the effects of an LMF. Accordingly, applying an LMF of varying magnitude, between $0~mG$ and 
$200~mG$, we observe uniform absorption across the whole beam, that increases with the magnitude of the TMF (Fig.~\ref{fig3}). The variation of transparency for the whole beam is shown in Fig. \ref{fig3} (g), displaying the same Hanle-EIT profile as Fig. \ref{fig2} (i). Such results are adequately described by the Zeeman effect and are no different to prior experiments that rely on linearly polarized light.

\begin{figure}[ht!]
\centering
\includegraphics[width=\linewidth]{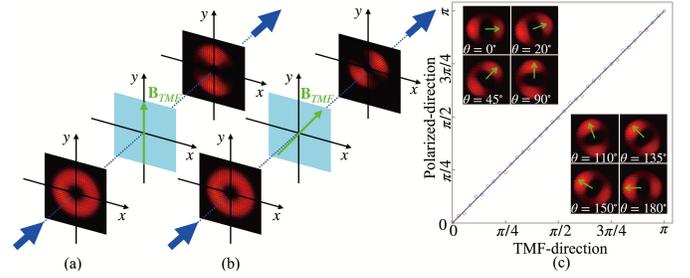}
\caption{Transmission profiles as function of TMFs alignment. (a) and (b): intensity and polarization distributions for vertical and diagonal TMF alignment. (c) Image axis of the transmission profile as a function of TMF alignment. Insets: examples of observed transmission profiles.}
\label{fig4}
\end{figure}

The transmitted vector beam, as a whole, is highly sensitive to the TMF however, particularly in regard to the $B$-field's alignment. To characterize the axis of the TMF, we select a radially polarized VB which generates the two-petal pattern after passing through the vapor cell. We then define the {\it image axis} as the line that passes through the maximal transmission regions of the two petals and the beam center. As mentioned previously, the visibility of the transmission profile can be controlled by the magnitude of the TMF, with a stronger magnetic field corresponding to stronger maximal absorption. We set the TMF to $230~mG$ to ensure maximal contrast, allowing us to identify the image axis as precisely as possible. We further note that the linear polarization of the transmitted region is also parallel to the image axis, providing an alternative way to identify the TMF axis. 

Fig. \ref{fig4} (a) and (b) show the relationship between the TMF alignment and the transmitted polarization, as reconstructed from CCD measurements of the Stokes parameters. Fig. \ref{fig4} (c) shows the angle of the image axis (green arrow) when rotating the TMF axis from $0$ to $\pi~$rad. Moreover, the VBs' polarization can also be manipulated by rotating the half wave plate and the VRP, producing the same rotational results as when the angle of the TMF axis is fixed. The axis of TMF can thus be easily observed and the TMF's strength can be measured similarly to the procedure outlined in Fig. \ref{fig2}.

\begin{figure}[ht!]
\centering
\includegraphics[width=\linewidth]{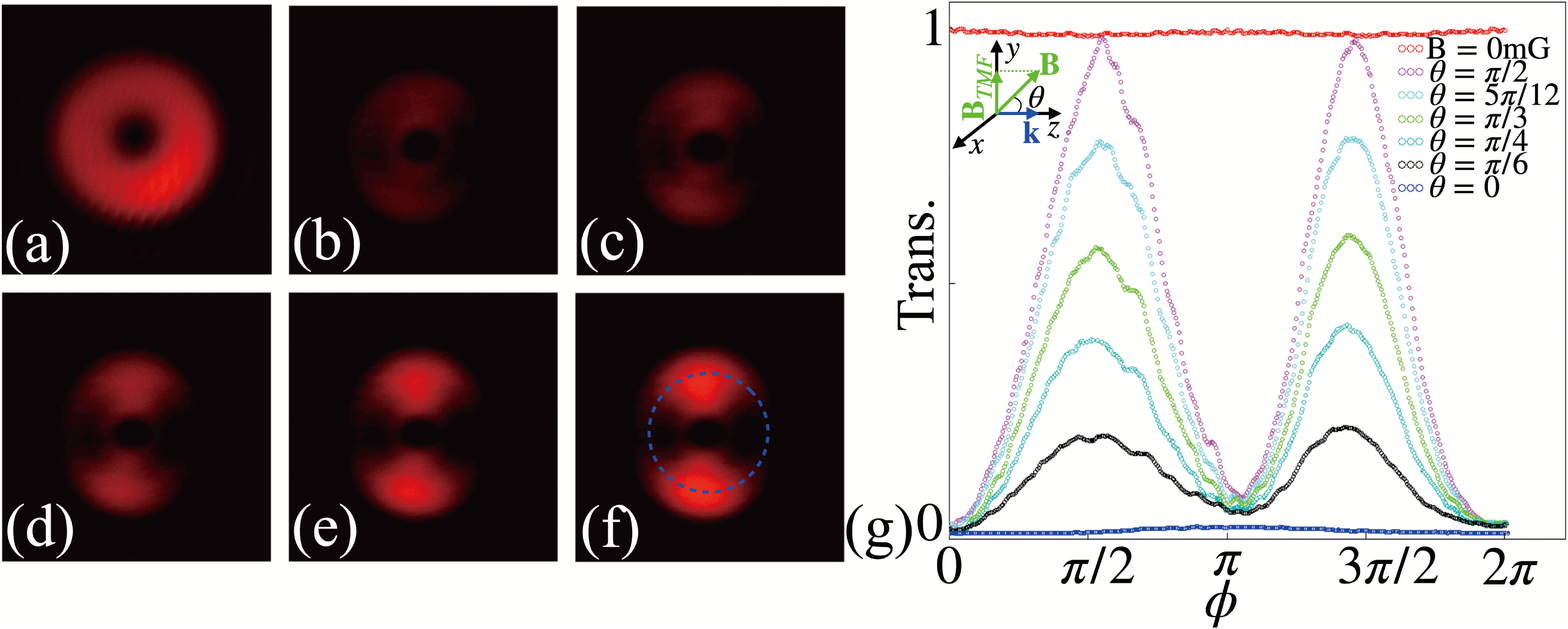}
\caption{The experimental results of the radially polarized beam in presence of the spatial magnetic field with fixed strength ($\left | \mathbf{B} \right |=230mG$). (a): $\left | \mathbf{B} \right |=0mG$. (b) - (f): transmitted patterns with $\theta = \pi/6, \pi/4, \pi/3, 5\pi/12, \pi/2$, respectively. (g) Polar plots for patterns at different angle $\theta$ at the radius indicated in (f).}
\label{fig5}
\end{figure}

To visualise arbitrary magnetic field alignments, we can combine our observations for a TMF and an LMF with further experiments that consider an arbitrary inclination angle, $\theta$, between the $\mathbf{B}$-field and the propagation axis, $\mathbf{k}$. For a TMF along the vertical axis, $\theta$ then denotes the angle in the $y-z$ plane and the magnetic field can be written as $\mathbf{B}=\mathbf{|B|} (0, \sin{\theta}, \cos{\theta})^T$, as reported in Fig. \ref{fig5}. 
In the absence of a magnetic field, there is no petal-like pattern and the transmission of the radially polarized VB is uniformly distributed along the azimuthal angle, similarly to Fig. \ref{fig2} (b) and Fig. \ref{fig3} (a).

We then set the strength of the magnetic field to $\mathbf{|B|}$ = 230 mG. When $\theta = 0$, the magnetic field is aligned to $\mathbf{k}$ , corresponding to a pure LMF, destroying Hanle resonances and resulting in strong homogeneous absorption as discussed in the context of Fig. \ref{fig3}. By increasing $\theta$, the petal-like pattern gradually appears according to the strength of the TMF component. In the case of $\theta = \pi/2$, the magnetic field is purely transverse, and the transmission profile is the same as for Fig. \ref{fig4}. As expected, regions where the polarization is perpendicular to the $\mathbf{B}$-$\mathbf{k}$ plane experience maximal absorption. However, with larger angle $\theta$, the TMF component of $\mathbf{B}$ increases and induces transparency in regions with parallel polarization. The relationship and sensitivity of transmitted patterns to the angle $\theta$ are captured in Fig. \ref{fig5} (g). The visibility of the pattern also depends on the magnitude of the field. A rotation of the magnetic field around the azimuthal angle $\phi$ would result in a corresponding shift of the absorption pattern. Visual inspection of the absorption pattern therefore gives maximal information on the magnetic field alignment, subject to the symmetry of the probe pattern.

\begin{figure}[ht!]
\centering
\includegraphics[width=\linewidth]{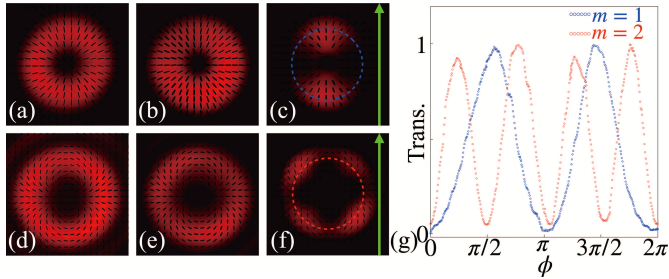}
\caption{Polarization and intensity profiles for VBs with different polarization topological charges with  $m$ = 1 (top row) and $m$ = 2 (bottom row). (a) and (d) profiles of VBs without atoms, (b) and (e) after passing through atoms in the absence of a magnetic field, (c) and (f) petal-like patterns under $\mathbf{B}_{\rm TMF}$ = 230 mG. (g) Polar plots of the absorption profile in (c) along the radius of largest contrast.}
\label{fig6}
\end{figure}

So far, we have considered only radially polarized VBs, but similar observations hold for general VBs. We demonstrate this in Fig. \ref{fig6} by comparing VBs with differing topological charge: $m$ = 1 and $m$ = 2.  
Fig. \ref{fig6} (a) and (d) show the polarization and intensity profile of the generated VBs with without atoms, respectively. These donut-like profiles change very little without appropriate shielding from magnetic fields, as shown in Fig. \ref{fig6} (b) and (e).  However, in the presence of a TMF, the transmitted beams show a two-fold and four-fold petal pattern: as shown in Fig. \ref{fig6} (c) and (f). Polar plots of the absorption profile with atoms, for $m$ = 1 and $m$ = 2, are shown in Fig. \ref{fig6} (g) and are in agreement with the original observation in cold atoms \cite{radwell2015spatially}.

\section{Theoretical interpretation and discussion}

As is now well known, atom-light interaction is strongly polarization dependent. There are however, infinitely many ways to decompose a field's polarization and to choose an atom's quantization axis. For the former, the spherical-basis has many useful properties \cite{yudin2010vector,lee1998sensitive,auzinsh2010optically}.  Here, light polarized perpendicularly to the atom's quantization axis drives a superposition of $\sigma^{+}$ and $\sigma^{-}$ transitions ($\Delta m_{\rm F} = \pm1$), while light polarized parallel to the quantization axis drives the associated $\pi$ transition ($\Delta m_{\rm F} = 0$). In the absence of a magnetic field, it is convenient to choose the quantization axis, $z$, along the propagation axis, $\mathbf{k}$, so that any light, $\mathbf{E}$,  polarized in the $x-y$ plane is simply formed from the superposition of two orthogonal circular components with equal amplitude and a varying phase difference. In the presence of a magnetic field however, it is helpful to choose the quantization axis along the axis of the magnetic field, $\mathbf{B}$,  so that the interaction is only dependent on the angle between $\mathbf{B}$ and $\mathbf{E}$.

Thus, we define the optical field in the spherical basis$\left \{ \mathbf{e_{0}=e_{z}, e_{\pm1} = \mp (e_{x}\pm \textrm{i} e_{y})/\sqrt{2}} \right \}$ \cite{yudin2010vector}:
\begin{equation}
\begin{aligned}
\mathbf{E} 
&=E_{0}\left((\cos{\alpha})\mathbf{e_{0}}+\frac{\sin{\alpha}}{\sqrt{2}}(-{\rm e}^{-i \beta_{1}}\mathbf{e_{+1}}+{\rm e}^{+i \beta_{2}}\mathbf{e_{-1}})\right),
\label{2}
\end{aligned}
\end{equation}
where $\mathbf{e}_{i}~\forall i \in \{x,y,z\}$, are the basis unit vectors in Cartesian coordinates and $\mathbf{e}_{j}~\forall j \in \{0,+1,-1\}$ represent the spherical basis with the quantization axis set by the magnetic field. Here, $\pi$-polarized, left and right circularly polarized light corresponds to $\mathbf{e}_{0,+1,-1}$ respectively, $E_{0}$ is the amplitude of light, $\alpha$ is the angle between $\mathbf{B}$ and $\mathbf{E}$ and $\beta_{1}$ and $\beta_{2}$ are the phase of each circular polarization. 

\begin{figure}[ht!]
\centering
\includegraphics[width=\linewidth]{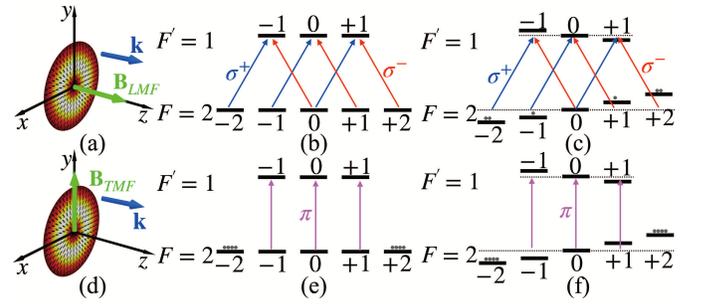}
\caption{Excitation scheme for the LMF and the TMF. Coherent dark state (b) and decoherent state (c) induced by Zeeman splitting. Bare dark state without (e) and with (f) Zeeman splitting. In presence of the magnetic field,
magnetic sublevels are shifted by an amount $\mu_{\rm B}g_{\rm F}m_{\rm F}B$, where $\mu_{\rm B}$ is the Bohr magneton, $g_{\rm F}$ is the Land$\acute{e}$-factor, and $B$ is the magnetic field strength.}
\label{fig7}
\end{figure}

\subsection{Interaction under an LMF}

As shown in Fig. \ref{fig7} (a), when an LMF is applied, we choose the quantization axis  along the LMF, coinciding with the propagation direction of the light. In this case, all the linearly polarized components of the probe VBs are perpendicular to the quantization axis, and the light will connect all Zeeman sublevels of the $F = 2 \to F^{'} = 1$ transition via multiple $\Lambda$ schemes with simultaneous $\sigma_{\pm1}$ excitations \cite{cox2011measurements,dancheva2000coherent,meshulam2007transfer}. According to the light polarization in Eq. (\ref{2}), the VBs shown in Eq. (\ref{1}) can be rewritten as:
\begin{equation}
\mathbf{E} = \frac{E_{0}}{\sqrt{2}} (\cos{(m\phi)} + \sin{(m\phi)})(-{\rm e}^{-i \beta_{1}(\phi)}\mathbf{e_{+1}} + {\rm e}^{+i \beta_{2}(\phi)}\mathbf{e_{-1}}),
\label{3}
\end{equation}
where, $\beta_{1}$ and $\beta_{2}$ are dependent on the azimuth.

When the LMF is zero, the Zeeman sublevels are degenerate, and atoms are pumped into a non-absorbing state induced by coherent population trapping. This is similar to the case of standard EIT, where the left and right circularly polarized components of an optical field resonate with the atomic levels to form the $\varLambda$ structure shown in Fig. \ref{fig7} (b). Here, a dark state due to coherent superposition of atomic energy levels is formed, which causes transparency of the whole VBs' profile. Increasing the strength of the LMF results in splitting of the Zeeman sublevels and an effective detuning as shown in Fig. \ref{fig7} (c). Thus, the coherent dark state is destroyed, the atoms can now interact and this leads to absorption of the probe VBs. For any linear polarization exciting the $\sigma_{\pm}$ transitions, the transparency is sensitive to the magnitude of the magnetic field and shows the Hanle-EIT profile \cite{anupriya2010hanle}.

\subsection{Interaction under a TMF}

When a pure TMF is applied, as shown in Fig. \ref{fig7} (d), the quantization axis is chosen to be aligned with the axis of the TMF. In contrast to the former case, the interaction of the linearly polarized components of the probe VBs is strongly dependent on the azimuthal angle. The components whose linear polarization is parallel to the TMF axis operate on the $\pi$ transition \cite{happer1972optical,huss2006polarization,yin2016tunable}, while the orthogonal components activate the $\sigma_{\pm}$ transitions. Other linearly polarized components can be considered as superpositions of these special cases and the inclined angle between the $\mathbf{B}_{\rm TMF}$ and the $\mathbf{E}$ determines which transition is dominant. Here, the situation of perpendicular components is similar to interaction under the $\mathbf{B}_{\rm TMF}$ and shows the same sensitivity to the magnetic field strength. Assuming the $\mathbf{B}_{\rm TMF}$ is along the $y$ axis and combining Eq. (\ref{1}) and Eq. (\ref{2}), then the $\mathbf{E}$-field of the VBs can be rewritten as:
\begin{equation}
\begin{aligned}
\mathbf{E}&={E_{0}}\sin{(m\phi)}\mathbf{e_{0}}+\frac{E_{0}}{\sqrt{2}}\cos{(m\phi)}(-{\rm e}^{-i \beta_{1}}\mathbf{e_{+1}}+{\rm e}^{+i \beta_{2}}\mathbf{e_{-1}})\\
&=E_{0}\sin{(m\phi)}\mathbf{e_{\parallel}}+E_{0}\cos{(m\phi)}\mathbf{e_{\perp}},
\label{4}
\end{aligned}
\end{equation}
where $\mathbf{e_{\parallel}}=\mathbf{e_{0}}$ and $\mathbf{e_{\perp}}=(-{\rm e}^{-i \beta_{1}}\mathbf{e_{+1}}+{\rm e}^{+i \beta_{2}}\mathbf{e_{-1}})/\sqrt{2}$.
The first term in Eq. (\ref{4}) is the linear component with polarization direction along the TMF axis driving $\pi$ transitions. In this situation, atoms are optically pumped into the stretch states by means of spontaneous emission and removed from the optical transition. Here, another type of dark state, due to strong optical pumping, is formed which will cause transparency of parallel components and shows insensitivity to the magnetic field (Fig. \ref{fig7} (e) and (f)). The second term in Eq. (\ref{4}) is the linear component with polarization perpendicular to the TMF axis. In this position, the analysis is similar to the interaction under the $\mathbf{B}_{\rm TMF}$ as discussed before. Increasing $\mathbf{B}_{\rm TMF}$ enhances the absorption (destroying the coherence) of perpendicular components and the splitting of the beam profile, since parallel components are always transmitted. Thus, the transmission profile of the probe VBs under the non-zero TMF fellows the equation as:
\begin{equation}
I\propto\left |\sin{(m\phi)}\right |^{2},
\label{5}
\end{equation}
satisfying the $2m$ sinusoidal transmission profile observed in Fig. \ref{fig6} (g).

\subsection{Interaction with arbitrarily oriented B-fields and associated applications}

Generally, when $\mathbf{B}$ is not applied along the axis of light propagation, it can be decomposed into contributions of an LMF and a TMF. Thus, for the general case, the interaction of the light with atoms can be viewed as a combination of the cases discussed in sections $\mathbf{A}$ and  $\mathbf{B}$, including all three transitions. A coherent dark state only appears when $\mathbf{B}=0$ and forms from pure $\sigma_{+}$ and $\sigma_{-}$ transitions. Frequency detunings (Zeeman splitting) and $\pi$ transitions induced by the magnetic field break the coherence of these dark states. Although the $\pi$ transition associated with the TMF induces bare dark states which are insensitive to the magnetic field.

Based on the above analysis, the transmitted pattern of VBs after passing through the atoms is strongly dependent on the $\mathbf{B}$, making this configuration a potentially useful tool for exploring spatially varying magnetic fields. The radially polarized beam whose polarization distribution resembles a compass would be used as the probe and the transmitted pattern has two petals whose orientation clearly show the axis of the TMF. Besides, by comparing maximum intensity of the transmitted pattern with initial intensity of the beam in the same position, the angle between the magnetic field and the plane of the beam profile can be easily obtained as shown in Fig. \ref{fig5} (g). By combining the included angle with the axis of TMF, the axis of the spatial $\mathbf{B}$ could be defined. Not only the alignment of the magnetic field, but also its strength influence the observed transmission pattern. The strength of the magnetic field  is associated with the intensity of the region perpendicular to two petals, since the polarization of this part is always perpendicular to the quantization axis set by the $\mathbf{B}$. The transparency of this region depends on the Zeeman splitting of atoms influenced by the magnetic field, which means the magnitude of the magnetic field can be deduced by measuring the intensity of this region.

Ultimately, the alignmnt and strength of a spatial $\mathbf{B}$-field can be seen and quantified from the transmitted vector beam. There are experimental limitations however. First of all, the measurement range of the magnetic field is limited by the atomic coherence. In this experiment, the Zeeman energy levels were used to build the atomic coherence, but hyperfine levels could expand the range of measurement of this configuration. Secondly, the direction of the spatial $\mathbf{B}$-field can not be obtained here. Since only resonant light is used in this experiment, the frequency detuning induced by spatial $\mathbf{B}$ has the same influence on both $\sigma_{+}$ and $\sigma_{-}$ transitions. After passing through atoms, the linear polarization can always be obtained which indicates two circular components have equal amplitudes and experience same absorption. Further study will be carried out to solve this problem by utilizing detuned light and measuring the polarization ellipse of transmitted parts \cite{selyem2019three, clark2016sculpting}.

\section{Conclusion}

In conclusion, we have investigated the transmission properties and pattern formation of vector beams in an atomic vapor, as influenced by a magnetic field. In particular, there are two limiting cases: corresponding to two kinds of dark state. When an LMF is applied, an incoming probe beam undergoes uniform absorption, and the coherence between Zeeman sublevels (coherent dark states) can be destroyed by increasing the strength of the LMF. Applying a TMF however, will generate bare dark states and produce a petal-like pattern which is dependent on the azimuthal angle and topological charge of the polarization. The general case, where the magnetic field is applied along an arbitrary axis, is also studied: revealing the general influence of a spatial $\mathbf{B}$-field on transmitted patterns. Thus, the information about both the alignment and the strength of a spatial $\mathbf{B}$-field can be seen in the transmitted pattern of a vector beam, providing a powerful tool in the investigation of 3D magnetic fields. Recent works on chip-scale VBs generation \cite{chen2020vector} and atomic components \cite{garrido2019compact,stern2019chip,mcgilligan2020laser} would be an exciting next step for realizing miniaturization. We also note that similar effects are seen in related work in the diamond (nitrogen-vacancy center) \cite{chen2020calibration} and cold atoms \cite{castellucci2021atomic} carried out, confirming the suitability of VBs for visual observation of the alignment of magnetic fields in 3D space.

\section*{Funding} This work was supported by the National Natural Science Foundation of China (92050103, 11774286, 11534008, 11604257 and 11574247) and the Fundamental Research Funds for the Central Universities of China. FC and SF-A acknowledge financial support
from the European Training Network ColOpt, which is funded by the European Union (EU) Horizon 2020 program
under the Marie Sklodowska-Curie Action, Grant Agreement No. 721465.
TWC acknowledges support by the National Research, Development and Innovation Office of Hungary (NKFIH) within the Quantum Technology National Excellence
Program (Project No. 2017-1.2.1-NKP-2017-00001). 

\section*{Disclosures} The authors declare no conflicts of interest.

\bibliography{reference}

\begin{thebibliography}{10}
\newcommand{\enquote}[1]{``#1''}

\bibitem{boller1991observation}
K.-J. Boller, A.~Imamo{\u{g}}lu, and S.~E. Harris, \enquote{Observation of
  electromagnetically induced transparency,} {\protect\JournalTitle{Physical
  Review Letters}} \textbf{66}, 2593 (1991).

\bibitem{fulton1995effects}
D.~J. Fulton, R.~R. Moseley, S.~Shepherd, B.~D. Sinclair, and M.~H. Dunn,
  \enquote{Effects of zeeman splitting on electromagnetically-induced
  transparency,} {\protect\JournalTitle{Optics Communications}} \textbf{116},
  231--239 (1995).

\bibitem{alezama1999eia}
S.~A.Lezama and A.M.Akulshin, \enquote{electromagnetically induced absorption,}
  {\protect\JournalTitle{Physical Review A}} \textbf{59}, 4732 (1999).

\bibitem{renzoni1997coherent}
F.~Renzoni, W.~Maichen, L.~Windholz, and E.~Arimondo, \enquote{Coherent
  population trapping with losses observed on the hanle effect of the d1 sodium
  line,} {\protect\JournalTitle{Physical Review A}} \textbf{55}, 3710 (1997).

\bibitem{kitching2011atomic}
J.~Kitching, S.~Knappe, and E.~A. Donley, \enquote{Atomic sensors--a review,}
  {\protect\JournalTitle{IEEE Sensors Journal}} \textbf{11}, 1749--1758 (2011).

\bibitem{wiesendanger2011single}
R.~Wiesendanger, \enquote{Single-atom magnetometry,}
  {\protect\JournalTitle{Current Opinion in Solid State and Materials Science}}
  \textbf{15}, 1--7 (2011).

\bibitem{drung1990low}
D.~Drung, R.~Cantor, M.~Peters, H.~Scheer, and H.~Koch, \enquote{Low-noise
  high-speed dc superconducting quantum interference device magnetometer with
  simplified feedback electronics,} {\protect\JournalTitle{Applied Physics
  Letters}} \textbf{57}, 406--408 (1990).

\bibitem{pannetier2004femtotesla}
M.~Pannetier, C.~Fermon, G.~Le~Goff, J.~Simola, and E.~Kerr,
  \enquote{Femtotesla magnetic field measurement with magnetoresistive
  sensors,} {\protect\JournalTitle{Science}} \textbf{304}, 1648--1650 (2004).

\bibitem{kominis2003subfemtotesla}
I.~Kominis, T.~Kornack, J.~Allred, and M.~V. Romalis, \enquote{A subfemtotesla
  multichannel atomic magnetometer,} {\protect\JournalTitle{Nature}}
  \textbf{422}, 596--599 (2003).

\bibitem{fleischhauer1994quantum}
M.~Fleischhauer and M.~O. Scully, \enquote{Quantum sensitivity limits of an
  optical magnetometer based on atomic phase coherence,}
  {\protect\JournalTitle{Physical Review A}} \textbf{49}, 1973 (1994).

\bibitem{alipieva2003narrow}
E.~Alipieva, S.~Gateva, E.~Taskova, and S.~Cartaleva, \enquote{Narrow structure
  in the coherent population trapping resonance in rubidium,}
  {\protect\JournalTitle{Optics Letters}} \textbf{28}, 1817--1819 (2003).

\bibitem{gateva2007shape}
S.~Gateva, L.~Petrov, E.~Alipieva, G.~Todorov, V.~Domelunksen, and
  V.~Polischuk, \enquote{Shape of the coherent-population-trapping resonances
  and high-rank polarization moments,} {\protect\JournalTitle{Physical Review
  A}} \textbf{76}, 025401 (2007).

\bibitem{acosta2006nonlinear}
V.~Acosta, M.~Ledbetter, S.~Rochester, D.~Budker, D.~J. Kimball, D.~Hovde,
  W.~Gawlik, S.~Pustelny, J.~Zachorowski, and V.~Yashchuk, \enquote{Nonlinear
  magneto-optical rotation with frequency-modulated light in the geophysical
  field range,} {\protect\JournalTitle{Physical Review A}} \textbf{73}, 053404
  (2006).

\bibitem{afach2015highly}
S.~Afach, G.~Ban, G.~Bison, K.~Bodek, Z.~Chowdhuri, Z.~D. Gruji\'{c}, L.~Hayen,
  V.~H\'{e}laine, M.~Kasprzak, K.~Kirch, P.~Knowles, H.-C. Koch, S.~Komposch,
  A.~Kozela, J.~Krempel, B.~Lauss, T.~Lefort, Y.~Lemi\`{e}re,
  A.~Mtchedlishvili, O.~Naviliat-Cuncic, F.~M. Piegsa, P.~N. Prashanth,
  G.~Qu\'{e}m\'{e}ner, M.~Rawlik, D.~Ries, S.~Roccia, D.~Rozpedzik,
  P.~Schmidt-Wellenburg, N.~Severjins, A.~Weis, E.~Wursten, G.~Wyszynski,
  J.~Zejma, and G.~Zsigmond, \enquote{Highly stable atomic vector magnetometer
  based on free spin precession,} {\protect\JournalTitle{Optics Express}}
  \textbf{23}, 22108--22115 (2015).

\bibitem{bison2018sensitive}
G.~Bison, V.~Bondar, P.~Schmidt-Wellenburg, A.~Schnabel, and J.~Voigt,
  \enquote{Sensitive and stable vector magnetometer for operation in zero and
  finite fields,} {\protect\JournalTitle{Optics Express}} \textbf{26},
  17350--17359 (2018).

\bibitem{zhang2019multi}
G.~Zhang, S.~Huang, F.~Xu, Z.~Hu, and Q.~Lin, \enquote{Multi-channel spin
  exchange relaxation free magnetometer towards two-dimensional vector
  magnetoencephalography,} {\protect\JournalTitle{Optics Express}} \textbf{27},
  597--607 (2019).

\bibitem{novikova2001compensation}
I.~Novikova, A.~Matsko, V.~Velichansky, M.~O. Scully, and G.~R. Welch,
  \enquote{Compensation of ac stark shifts in optical magnetometry,}
  {\protect\JournalTitle{Physical Review A}} \textbf{63}, 063802 (2001).

\bibitem{pustelny2006pump}
S.~Pustelny, D.~J. Kimball, S.~Rochester, V.~Yashchuk, W.~Gawlik, and
  D.~Budker, \enquote{Pump-probe nonlinear magneto-optical rotation with
  frequency-modulated light,} {\protect\JournalTitle{Physical Review A}}
  \textbf{73}, 023817 (2006).

\bibitem{budker1998nonlinear}
D.~Budker, V.~Yashchuk, and M.~Zolotorev, \enquote{Nonlinear magneto-optic
  effects with ultranarrow widths,} {\protect\JournalTitle{Physical Review
  Letters}} \textbf{81}, 5788 (1998).

\bibitem{budker2000sensitive}
D.~Budker, D.~Kimball, S.~Rochester, V.~Yashchuk, and M.~Zolotorev,
  \enquote{Sensitive magnetometry based on nonlinear magneto-optical rotation,}
  {\protect\JournalTitle{Physical Review A}} \textbf{62}, 043403 (2000).

\bibitem{novikova2000ac}
I.~Novikova, A.~Matsko, V.~Sautenkov, V.~Velichansky, G.~Welch, and M.~Scully,
  \enquote{Ac-stark shifts in the nonlinear faraday effect,}
  {\protect\JournalTitle{Optics Letters}} \textbf{25}, 1651--1653 (2000).

\bibitem{pustelny2008magnetometry}
S.~Pustelny, A.~Wojciechowski, M.~Gring, M.~Kotyrba, J.~Zachorowski, and
  W.~Gawlik, \enquote{Magnetometry based on nonlinear magneto-optical rotation
  with amplitude-modulated light,} {\protect\JournalTitle{Journal of Applied
  Physics}} \textbf{103}, 063108 (2008).

\bibitem{shah2007subpicotesla}
V.~Shah, S.~Knappe, P.~D. Schwindt, and J.~Kitching, \enquote{Subpicotesla
  atomic magnetometry with a microfabricated vapour cell,}
  {\protect\JournalTitle{Nature Photonics}} \textbf{1}, 649--652 (2007).

\bibitem{budker2007optical}
D.~Budker and M.~Romalis, \enquote{Optical magnetometry,}
  {\protect\JournalTitle{Nature Physics}} \textbf{3}, 227--234 (2007).

\bibitem{le2013optical}
D.~Le~Sage, K.~Arai, D.~R. Glenn, S.~J. DeVience, L.~M. Pham, L.~Rahn-Lee,
  M.~D. Lukin, A.~Yacoby, A.~Komeili, and R.~L. Walsworth, \enquote{Optical
  magnetic imaging of living cells,} {\protect\JournalTitle{Nature}}
  \textbf{496}, 486--489 (2013).

\bibitem{lee1998sensitive}
H.~Lee, M.~Fleischhauer, and M.~O. Scully, \enquote{Sensitive detection of
  magnetic fields including their orientation with a magnetometer based on
  atomic phase coherence,} {\protect\JournalTitle{Physical Review A}}
  \textbf{58}, 2587 (1998).

\bibitem{dimitrijevic2008role}
J.~Dimitrijevi{\'c}, A.~Krmpot, M.~Mijailovi{\'c}, D.~Arsenovi{\'c},
  B.~Pani{\'c}, Z.~Gruji{\'c}, and B.~Jelenkovi{\'c}, \enquote{Role of
  transverse magnetic fields in electromagnetically induced absorption for
  elliptically polarized light,} {\protect\JournalTitle{Physical Review A}}
  \textbf{77}, 013814 (2008).

\bibitem{yudin2010vector}
V.~Yudin, A.~Taichenachev, Y.~Dudin, V.~Velichansky, A.~Zibrov, and S.~Zibrov,
  \enquote{Vector magnetometry based on electromagnetically induced
  transparency in linearly polarized light,} {\protect\JournalTitle{Physical
  Review A}} \textbf{82}, 033807 (2010).

\bibitem{cox2011measurements}
K.~Cox, V.~I. Yudin, A.~V. Taichenachev, I.~Novikova, and E.~E. Mikhailov,
  \enquote{Measurements of the magnetic field vector using multiple
  electromagnetically induced transparency resonances in rb vapor,}
  {\protect\JournalTitle{Physical Review A}} \textbf{83}, 015801 (2011).

\bibitem{margalit2013degenerate}
L.~Margalit, M.~Rosenbluh, and A.~Wilson-Gordon, \enquote{Degenerate two-level
  system in the presence of a transverse magnetic field,}
  {\protect\JournalTitle{Physical Review A}} \textbf{87}, 033808 (2013).

\bibitem{zhan2009cylindrical}
Q.~Zhan, \enquote{Cylindrical vector beams: from mathematical concepts to
  applications,} {\protect\JournalTitle{Advances in Optics and Photonics}}
  \textbf{1}, 1--57 (2009).

\bibitem{wang2020vectorial}
J.~Wang, F.~Castellucci, and S.~Franke-Arnold, \enquote{Vectorial light--matter
  interaction: Exploring spatially structured complex light fields,}
  {\protect\JournalTitle{AVS Quantum Science}} \textbf{2}, 031702 (2020).

\bibitem{fatemi2011cylindrical}
F.~K. Fatemi, \enquote{Cylindrical vector beams for rapid
  polarization-dependent measurements in atomic systems,}
  {\protect\JournalTitle{Optics Express}} \textbf{19}, 25143--25150 (2011).

\bibitem{wang2018optically}
J.~Wang, X.~Yang, Y.~Li, Y.~Chen, M.~Cao, D.~Wei, H.~Gao, and F.~Li,
  \enquote{Optically spatial information selection with hybridly polarized beam
  in atomic vapor,} {\protect\JournalTitle{Photonics Research}} \textbf{6},
  451--456 (2018).

\bibitem{yang2019manipulating}
X.~Yang, A.~Fang, J.~Wang, Y.~Li, X.~Chen, X.~Zhang, M.~Cao, D.~Wei,
  K.~M\"{u}ller-Dethlefs, H.~Gao, and F.~Li, \enquote{Manipulating the
  transmission of vector beam with spatially polarized atomic ensemble,}
  {\protect\JournalTitle{Optics Express}} \textbf{27}, 3900--3908 (2019).

\bibitem{wang2019directly}
J.~Wang, X.~Yang, Z.~Dou, S.~Qiu, J.~Liu, Y.~Chen, M.~Cao, H.~Chen, D.~Wei,
  K.~M{\"u}ller-Dethlefs, H.~Gao, and F.~Li, \enquote{Directly extracting the
  authentic basis of cylindrical vector beams by a pump-probe technique in an
  atomic vapor,} {\protect\JournalTitle{Applied Physics Letters}} \textbf{115},
  221101 (2019).

\bibitem{wang2020optically}
J.~Wang, Y.~Chen, X.~Yang, J.~Liu, S.~Qiu, M.~Cao, H.~Chen, D.~Wei,
  K.~M{\"u}ller-Dethlefs, H.~Gao, and F.~Li, \enquote{Optically polarized
  selection in atomic vapor and its application in mapping the polarization
  distribution,} {\protect\JournalTitle{Journal of Physics Communications}}
  \textbf{4}, 015019 (2020).

\bibitem{shi2015magnetic}
S.~Shi, D.-S. Ding, Z.-Y. Zhou, Y.~Li, W.~Zhang, and B.-S. Shi,
  \enquote{Magnetic-field-induced rotation of light with orbital angular
  momentum,} {\protect\JournalTitle{Applied Physics Letters}} \textbf{106},
  261110 (2015).

\bibitem{stern2016controlling}
L.~Stern, A.~Szapiro, E.~Talker, and U.~Levy, \enquote{Controlling the
  interactions of space-variant polarization beams with rubidium vapor using
  external magnetic fields,} {\protect\JournalTitle{Optics Express}}
  \textbf{24}, 4834--4841 (2016).

\bibitem{bouchard2016polarization}
F.~Bouchard, H.~Larocque, A.~M. Yao, C.~Travis, I.~De~Leon, A.~Rubano,
  E.~Karimi, G.-L. Oppo, and R.~W. Boyd, \enquote{Polarization shaping for
  control of nonlinear propagation,} {\protect\JournalTitle{Physical Review
  Letters}} \textbf{117}, 233903 (2016).

\bibitem{hu2019nonlinear}
H.~Hu, D.~Luo, and H.~Chen, \enquote{Nonlinear frequency conversion of vector
  beams with four wave mixing in atomic vapor,} {\protect\JournalTitle{Applied
  Physics Letters}} \textbf{115}, 211101 (2019).

\bibitem{parigi2015storage}
V.~Parigi, V.~D’Ambrosio, C.~Arnold, L.~Marrucci, F.~Sciarrino, and
  J.~Laurat, \enquote{Storage and retrieval of vector beams of light in a
  multiple-degree-of-freedom quantum memory,} {\protect\JournalTitle{Nature
  Communications}} \textbf{6}, 1--7 (2015).

\bibitem{ye2019experimental}
Y.-H. Ye, M.-X. Dong, Y.-C. Yu, D.-S. Ding, and B.-S. Shi,
  \enquote{Experimental realization of optical storage of vector beams of light
  in warm atomic vapor,} {\protect\JournalTitle{Optics Letters}} \textbf{44},
  1528--1531 (2019).

\bibitem{radwell2015spatially}
N.~Radwell, T.~W. Clark, B.~Piccirillo, S.~M. Barnett, and S.~Franke-Arnold,
  \enquote{Spatially dependent electromagnetically induced transparency,}
  {\protect\JournalTitle{Physical Review Letters}} \textbf{114}, 123603 (2015).

\bibitem{yang2019observing}
X.~Yang, Y.~Chen, J.~Wang, Z.~Dou, M.~Cao, D.~Wei, H.~Batelaan, H.~Gao, and
  F.~Li, \enquote{Observing quantum coherence induced transparency of hybrid
  vector beams in atomic vapor,} {\protect\JournalTitle{Optics Letters}}
  \textbf{44}, 2911--2914 (2019).

\bibitem{clark2016sculpting}
T.~W. Clark, \enquote{Sculpting shadows. on the spatial structuring of fields
  \& atoms: a tale of light and darkness,} Ph.D. thesis, University of Glasgow
  (2016).

\bibitem{castellucci2021atomic}
F.~Castellucci, T.~W. Clark, A.~Selyem, J.~Wang, and S.~Franke-Arnold,
  \enquote{An atomic compass--detecting 3d magnetic field alignment with vector
  vortex light,} {\protect\JournalTitle{arXiv preprint arXiv:2106.13360}}
  (2021).

\bibitem{marrucci2006optical}
L.~Marrucci, C.~Manzo, and D.~Paparo, \enquote{Optical spin-to-orbital angular
  momentum conversion in inhomogeneous anisotropic media,}
  {\protect\JournalTitle{Physical Review Letters}} \textbf{96}, 163905 (2006).

\bibitem{marrucci2006pancharatnam}
L.~Marrucci, C.~Manzo, and D.~Paparo, \enquote{Pancharatnam-berry phase optical
  elements for wave front shaping in the visible domain: switchable helical
  mode generation,} {\protect\JournalTitle{Applied Physics Letters}}
  \textbf{88}, 221102 (2006).

\bibitem{milione2011higher}
G.~Milione, H.~Sztul, D.~Nolan, and R.~Alfano, \enquote{Higher-order
  poincar{\'e} sphere, stokes parameters, and the angular momentum of light,}
  {\protect\JournalTitle{Physical Review Letters}} \textbf{107}, 053601 (2011).

\bibitem{auzinsh2010optically}
M.~Auzinsh, D.~Budker, and S.~Rochester, \emph{Optically polarized atoms:
  understanding light-atom interactions} (Oxford University Press, 2010).

\bibitem{dancheva2000coherent}
Y.~Dancheva, G.~Alzetta, S.~Cartaleva, M.~Taslakov, and C.~Andreeva,
  \enquote{Coherent effects on the zeeman sublevels of hyperfine states in
  optical pumping of rb by monomode diode laser,} {\protect\JournalTitle{Optics
  Communications}} \textbf{178}, 103--110 (2000).

\bibitem{meshulam2007transfer}
R.~Meshulam, T.~Zigdon, A.~Wilson-Gordon, and H.~Friedmann,
  \enquote{Transfer-of-coherence-enhanced stimulated emission and
  electromagnetically induced absorption in zeeman split f g--> f e= f g- 1
  atomic transitions,} {\protect\JournalTitle{Optics Letters}} \textbf{32},
  2318--2320 (2007).

\bibitem{anupriya2010hanle}
J.~Anupriya, N.~Ram, and M.~Pattabiraman, \enquote{Hanle electromagnetically
  induced transparency and absorption resonances with a laguerre gaussian
  beam,} {\protect\JournalTitle{Physical Review A}} \textbf{81}, 043804 (2010).

\bibitem{happer1972optical}
W.~Happer, \enquote{Optical pumping,} {\protect\JournalTitle{Reviews of Modern
  Physics}} \textbf{44}, 169 (1972).

\bibitem{huss2006polarization}
A.~Huss, R.~Lammegger, L.~Windholz, E.~Alipieva, S.~Gateva, L.~Petrov,
  E.~Taskova, and G.~Todorov, \enquote{Polarization-dependent sensitivity of
  level-crossing, coherent-population-trapping resonances to stray magnetic
  fields,} {\protect\JournalTitle{JOSA B}} \textbf{23}, 1729--1736 (2006).

\bibitem{yin2016tunable}
L.~Yin, B.~Luo, J.~Xiong, and H.~Guo, \enquote{Tunable rubidium excited state
  voigt atomic optical filter,} {\protect\JournalTitle{Optics Express}}
  \textbf{24}, 6088--6093 (2016).

\bibitem{selyem2019three}
A.~Selyem, \enquote{Three-dimensional light sculptures and their interaction
  with atomic media: an experimentalist's guide,} Ph.D. thesis, University of
  Glasgow (2019).

\bibitem{chen2020vector}
Y.~Chen, K.-Y. Xia, W.-G. Shen, J.~Gao, Z.-Q. Yan, Z.-Q. Jiao, J.-P. Dou,
  H.~Tang, Y.-Q. Lu, and X.-M. Jin, \enquote{Vector vortex beam emitter
  embedded in a photonic chip,} {\protect\JournalTitle{Physical Review
  Letters}} \textbf{124}, 153601 (2020).

\bibitem{garrido2019compact}
C.~L. Garrido~Alzar, \enquote{Compact chip-scale guided cold atom gyrometers
  for inertial navigation: Enabling technologies and design study,}
  {\protect\JournalTitle{AVS Quantum Science}} \textbf{1}, 014702 (2019).

\bibitem{stern2019chip}
L.~Stern, D.~G. Bopp, S.~A. Schima, V.~N. Maurice, and J.~E. Kitching,
  \enquote{Chip-scale atomic diffractive optical elements,}
  {\protect\JournalTitle{Nature Communications}} \textbf{10}, 1--7 (2019).

\bibitem{mcgilligan2020laser}
J.~P. Mcgilligan, K.~Moore, A.~Dellis, G.~Martinez, E.~de~Clercq, P.~Griffin,
  A.~Arnold, E.~Riis, R.~Boudot, and J.~Kitching, \enquote{Laser cooling in a
  chip-scale platform,} {\protect\JournalTitle{Applied Physics Letters}}
  \textbf{117}, 054001 (2020).

\bibitem{chen2020calibration}
B.~Chen, X.~Hou, F.~Ge, X.~Zhang, Y.~Ji, H.~Li, P.~Qian, Y.~Wang, N.~Xu, and
  J.~Du, \enquote{Calibration-free vector magnetometry using nitrogen-vacancy
  center in diamond integrated with optical vortex beam,}
  {\protect\JournalTitle{Nano Letters}} \textbf{20}, 8267--8272 (2020).

\end{thebibliography}

\end{document}